\documentclass[aps,amsmath,nofootinbib,superscriptaddress,amssymb,preprintnumbers,floatfix,showpacs,preprint]{revtex4-1}
 \usepackage{amsmath,txfonts,longtable,booktabs,overpic,amssymb,bm,bbm,multirow,float,graphicx,color,dcolumn,subfigure,hyperref,tikz,lineno}
 \usepackage{appendix}
 \hypersetup{colorlinks,citecolor=blue,anchorcolor=red,menucolor=red, linkcolor=red,filecolor=red,runcolor=red,urlcolor=blue,frenchlinks=true}
\begin{document}

\title{Repeating fast radio bursts produced by a strange star interacting with its planet in an eccentric orbit}

\author{Nurimangul Nurmamat}\email{nurimangul@smail.nju.edu.cn}
\affiliation{School of Astronomy and Space Science, Nanjing University, Nanjing 210023, China}

\author{Yong-Feng Huang \footnote{Correspondence Author. email: hyf@nju.edu.cn}}
\affiliation{School of Astronomy and Space Science, Nanjing University, Nanjing 210023, China}
\affiliation{Key Laboratory of Modern Astronomy and Astrophysics (Nanjing University), Ministry of Education, Nanjing 210023, China}
\affiliation{Xinjiang Astronomical Observatory, Chinese Academy of Sciences, Urumqi 830011, Xinjiang, China}

\author{Jin-Jun Geng}\email{gengjinjun@nju.edu.cn}
\affiliation{Purple Mountain Observatory, Chinese Academy of Sciences, Nanjing 210023, China}

\author{Abdusattar Kurban}\email{akurban@xao.ac.cn}
\affiliation{Xinjiang Astronomical Observatory, Chinese Academy of Sciences, Urumqi 830011, China}
\affiliation{Xinjiang Key Laboratory of Radio Astrophysics, Urumqi 830011, China}

\author{Bing Li}\email{libing@ihep.ac.cn}
\affiliation{Key Laboratory of Particle Astrophysics, Chinese Academy of Sciences, Beijing 100049, China}
\affiliation{Particle Astrophysics Division, Institute of High Energy Physics, Chinese Academy of Sciences, Beijing 100049, China}


\begin{abstract}
FRB 180916 is an important repeating fast radio burst (FRB)
source. Interestingly, the activity of
FRB 180916 shows a well-regulated behavior, with a period of 16.35
days. The bursts are found to occur in a duty cycle of about 5
days in each period. In this study, we suggest that the bursts of
FRB 180916 are produced by a strange star interacting with its
planet. The planet moves in a highly eccentric orbit around its
compact host, with the periastron only slightly beyond the tidal
disruption radius. As a result, the planet will be partially
disrupted every time it passes through the periastron. The
stripped material from the planet will be accreted by the strange
star, falling to the polar cap region along the magnetic field
lines and accumulated there. It will finally lead to a local
collapse when the crust at the polar region is overloaded,
triggering an FRB. The observed 16.35 day period corresponds to
the orbital motion of the planet, and the 5 day duty cycle is
explained as the duration of the partial disruption near the
periastron. The energy released in each local collapse event can be
as high as $\sim 10^{42}\,\rm {erg}$, which is large enough to account
for typical FRBs even if the radiation efficiency is low.
\end{abstract}
\maketitle
% \keywords{Radio transient sources; Exoplanets; Strange matter; strange star Tidal disruption}

\section{Introduction}
Strange quark matter (SQM), which contains almost equal number of
up, down and strange quarks, may be the true ground state of
hadrons~\cite{1984Witten}. If this SQM hypothesis is true, then
there should exist strange quark stars (also commonly shortened as
strange stars (SSs)). SSs are extremely compact stellar objects
that are mainly composed of SQM~\cite{1986Alcock,Haensel1986,Weber2005}. Comparing with
traditional neutron stars (NSs), an SS may have an even higher
mean density. SSs may be connected with various high energy
astronomical phenomena, such as millisecond magnetars, X-ray or
$\gamma$-ray bursters, Quark-nova, etc
\cite{1989Frieman,1997Bombaci,1998Cheng,Dai1998,2002Ouyed,2009Du}.
However, it is very difficult to distinguish between SSs and NSs
via observations. The main reason is that SSs may be covered by a
thin crust of normal hadronic matter \cite{1986Alcock}, which
makes SSs to be very much similar to NSs for an outside observer.
The total mass of an SS crust is believed to be in the range of
$10^{-7}$ --- $10^{-5}\,{M_{\odot}}$, with a thickness of $\sim
2\times 10^4\,\rm cm $. \cite{1997Huang,1997Huanga}
numerically calculated the mechanical equilibrium condition at the
bottom of the crust, and found that the maximum density at the
crust bottom should be much smaller than the so called neutron
drip density. As a result, a typical $1.4\,M_{\odot }$ SS cannot
have a crust more massive than $ M_{\rm crust} \sim 3.4 \times 10^{-6}
M_{\odot}$, with the crust thickness being less than $\sim 10^{4}\rm {cm}$
. In this case, when an SS accretes matter from the surrounding
medium, the crust will get heavier and heavier, which may finally
lead to the collapse of the crust in a very short time ($ \sim 5.4
\times 10^{-3}\,\rm {s}$), producing an intense explosion of short
duration. During the collapse, normal hadronic matter is converted
to SQM, and each baryon will release an energy of $\sim 6.3$ MeV.
The total energy will roughly be $10^{46}$ --- $10^{47} \rm erg$ \cite{2004Jia}.

Fast radio bursts (FRBs) are a new kind of high energy
astronomical phenomena characterized by a short duration ($\sim$ a
few milliseconds) and intense emission in radio waves
\cite{Lorimer2007Sci}. Their extra-galactic origin was firstly
hinted from the large dispersion measure (DM) along the line of
sight, and is later confirmed by direct redshift measurement in a
few cases \cite{Petroff2019AARv,Cordes2019ARAA,2020Natur}. It is
interesting to note that some FRB sources can burst out
repeatedly, even with periodicity being hinted in several of them.
Till now, nearly 800 different FRB sources have been detected,
among which more than 60 are repeaters \cite{2022Petroff,2022Xu,2022Lanman,2022Feng}
\footnote{\url{http://blinkverse.alkaidos.cn/\#/overview}}.
The host galaxies are also identified for 44 sources \cite{NG2023}
\footnote{\url{https://ecommons.cornell.edu/server/api/core/bitstreams/3c93d0f8-c307-4909-a9a0-26b7a2df1397/content }}
. However, the central
engine and trigger mechanism of FRBs are still largely uncertain.
It is widely believed that FRBs should be connected with
magnetars, and may be connected with some kinds of perturbations
in the magnetosphere \cite{2021Cui,2020MNRASLu}.
In addition to that, various other models have also been proposed and could
not be expelled yet \cite{Platts2019}. For example, motivated by the
atmosphere pollution of white dwarfs (WDs) by heavy
elements \cite{2015Vanderburg} and the WD-planet/asteroid tidal disruption
interactions \cite{2003Jura}, \cite{2022Kurban} proposed that periodically
repeating FRBs may be triggered by a magnetized NS interacting with its planet
in a highly eccentric orbit. The planet will be partially tidal disrupted
every time it comes to the periastron. The disrupted clumps then interact with
the pulsar wind and produce FRBs via the Alfv$\acute{\rm e}$n wing mechanism.

More interestingly, \cite{2018Zhang} argued that FRBs may be triggered by
the collapse of SS crust when the SS accretes matter from the surrounding
environment. In their study, the crust collapses as a whole to release a
huge amount of energy, but which also makes it difficult to produce FRBs
repeatedly. Recently, \cite{2021Geng} went further to argue that the
accreted matter will not diffuse on the crust due to the strong surface
magnetic field. As a result, only the crust at the polar cap region will
collapse when it is locally overloaded. In this way, an SS accreting from its
companion can naturally produce periodic FRBs, such as the repeating source
of FRB 180916. Note that in their framework, the periodicity is mainly due to
the thermal-viscous instability of the accretion disk. As a result, the
active window ($\sim 5$ days for FRB 180916) corresponds to a high accretion
state of the system, while the $\sim 16$ day period corresponds to the accumulation
timescale of the accretion disk. Both the active window and the period are
determined by the accretion rate and the viscosity parameter. Note that since
the periodicity comes from the thermal-viscous instability,
the period may not be stable and may vary in a relatively wide range.

In this study, the essential ingredients of  \cite{2022Kurban} and
\cite{2021Geng}'s models are incorporated to build a new model for periodically
repeating FRBs. We assume that a planet moves around a magnetized SS in a
highly eccentric orbit. The planet will be partially disrupted by tidal force
every time it comes across the periastron. The disrupted material is then
accreted by the SS, causing the polar cap to collapse and produce FRBs.
Note that the planet can be either a gaseous or a rocky one.
For a gaseous planet, it is easy to understand that the accreted material will
fall toward the polar regions of the SS along the magnetic field lines.
For a rocky planet, the partial disruption process occurs at a distance
of $\sim 10^{11} \rm {cm}$, producing rocky clumps a few kilometers in size.
Shearing strength then plays the major role to resist the tidal force for
these clumps. However, they will also be completely disrupted by the tidal
force when their distance to the SS is $\sim 10^{9} \rm cm$. Thus they will
still fall toward the polar regions along the magnetic field lines
\cite{2023MNRAS}.

The structure of our paper is organized as follows. In Section 2, the
basic features of highly eccentric planetary systems are introduced. The
collapsing process of the polar cap crust of an accreting SS is presented
in Section 3. In Section 4, the model is applied to explain the observed
repeating event of FRB 180916. Finally, Section 5 presents our conclusions
and some discussion.

\section{tidal disruption in a highly eccentric planetary system}

When an object gets too close to a compact star, it will be teared
apart and a tidal disruption event happens. In such a process, a
significant portion of the teared matter will be captured and
accreted by the compact star
\cite{1975NaturHills,1988Rees,2021Gezari}. In a tidal disruption
event, the compact star can be a BH, NS or WD, and the teared
object can be a main sequence star, a planet, an asteroid/comet,
or sometimes even a WD
\cite{2016vanVelzen,2022King,2013Niko,2003Jura,2015Vanderburg}.
Recently, \cite{2022Kurban} studied the interaction of a highly
magnetized NS with and its planet in a highly eccentric orbit.
They suggested that the planet will be partially disrupted when it
passes through the periastron. The disrupted clumps then interact
with the magnetar wind to give birth to FRBs.

In this study, we adopt the basic configuration of
\cite{2022Kurban}. The main difference is that we use a strange
star to replace the compact star of NS. The planet is still in a
highly eccentrical orbit. It will be partially disrupted by the SS
every time it passes through the periastron. The teared materials
are then accreted by the SS, falling toward the magnetic polar
regions. The accumulated materials will finally lead to a local
collapse of the SS crust at the polar region, producing an FRB.

We assume that the SS has a typical mass of $M_{\rm ss} = 1.4 M_{\odot}$.
The companion is a planet with the mass denoted as $m_{\rm {planet}}$
and the mean density denoted as $\bar{\rho}$.  The semimajor 
axis and period of its orbit are denoted as $a$ and $P_{\rm orb}$,
respectively. According to the Kepler's third law, the orbital
period of this planetary system can be expressed as
\begin{equation}
P_{\rm orb} =\left( \frac{4\pi^2 a^3}{G \left(M_{\rm ss} + m_{\rm planet}  \right)} \right)^\frac{1}{2},
\end{equation}
where $G$ is the gravitational constant. The separation ($R$) between
the SS and the planet in the eccentric orbit is related to the
eccentricity ($e$), semimajor axis ($a$) and phase ($\theta$) as
$R = a(1 - e^2)/ (1 + e\,\rm {cos}\,\theta)$, while the periastron
separation of the planet is $R_{\rm p} = a (1 - e)$.

The tidal disruption radius ($R_{\rm {T}}$), i.e., the separation at which
the planet will be completely disrupted by the tidal force of the
compact star \cite{1975NaturHills},
is $R_{\rm {T}} = \left( 6 M_{\rm {SS}} / \pi \bar{\rho}\right)^{1/3}$. When the separation is
slightly larger, i.e. $R_{\rm {T}} < R < 2.7\,R_{\rm{T}}$, the planet will be
partially disrupted \cite{Liu2013ApJ}. In the highly eccentric
orbit case considered here, the planet will be significantly
affected by the tidal force near the periastron, but will be
unaffected at other orbital phase. Note that if the periastron
is too close to the SS (i.e., with $R_{\rm {p}} \leq R_{\rm {T}}$), the planet
will be completely disrupted when it passes through the periastron.
On the other hand, if $R_{\rm {T}} < R_{\rm {p}} < 2.7 \,R_{\rm {T}}$, the planet will
only be partially disrupted every time it passes through the
periastron. In this study, we take $R_{\rm {p}} = 1.5 \,R_{\rm {T}}$ as a typical
case for the partial disruption configuration. Also, we only
consider the tidal interaction near the periastron for simplicity.
The condition of $R_{\rm {p}} = 1.5 \,R_{\rm {T}}$ then means
\begin{equation}
a\,(1-e) = 1.5\,\left( \frac{6 M_{\rm ss}}{\pi \bar{\rho }}\right)^{\frac{1}{3}}.
\end{equation}

Because our condition of $R_{\rm p} = 1.5\,R_{\rm T}$ satisfies the partial
disruption condition ($R_{\rm{T}} < R_{\rm {p}} < 2.7 \,R_{\rm{T}}$), there
exists a critical phase angle $\theta_{\rm {c}}$ defined by $R(\theta_{\rm {c}}) = 2.7 \,R_{\rm {T}}$.
When the planet is in the phase range of  $ - \theta_{\rm {c}} < \theta <
\theta_{\rm {c}}$, it will be seriously affected by tidal interaction and
the system will be in an active regime. Therefore, the active
period can be expressed as \cite{2022Kurban,2007Sepinsky}
\begin{equation}
   \Delta P_{\rm orb}=\frac{P_{\rm orb}}{2 \pi} \sqrt{(1-e^2)^3}
   \int_{-\theta_{\rm c}}^{\theta_{\rm c}}  \frac{1}{(1+e \cos\, \theta)^2} \,d{\theta}.
\end{equation}

The partial disruption is a very complicated process. It is difficult
to estimate how much material will be teared up from the planet and will
be finally accreted by the SS. It may depend on a lot of complicated
factors, such as the composition of the planet and its rotation. Here,
we roughly estimate the total accreted mass during one active period
as $M_{\rm {acc}} = \Delta P_{\rm {orb}}
\dot{M}_{\rm {Edd}}$, where $ \dot{M}_{\rm {Edd}}$
is the Eddington accretion rate
expressed as $\dot{M}_{\rm {Edd}} =
4 \pi r  m_{\rm p} c / \sigma_{\rm T} \cong 10^{18} \mathrm {g\,s^{-1}}$.
Here $r$ is the radius of the SS, $m_{\rm p}$ is the mass of proton,
$c$ is the speed of light, and $\sigma_{\rm T}$ is the cross section of
Thomson scattering.

\section{Collapse of the SS crust at polar region}

SSs are compact objects made up of almost equal numbers of up,
down and strange quarks. They may be covered by a normal matter
crust \cite{1986Alcock}. The maximum density at the bottom of
the crust should not exceed the definite limit of neutron drip
density ($\epsilon_{\rm {drip}}$), but is most likely
significantly less than that when the mechanical equilibrium is
considered \cite{1997Huang,1997Huanga}. As a result, the maximum
mass of the crust of a typical $1.4\,M_{\odot }$ SS is $M_{\rm {crust}}
\sim 3.4 \times 10^{-6}\, M_{\odot}$, with a thickness of $\sim
10^{4}\,\rm {cm}$.

When an SS accretes material from the surrounding medium, the
crust will be heavier and heavier and may finally collapse in a
very short time ($\sim 10^{-3}\,\rm {s}$). During the collapse, every
baryon may release an energy of about 6.3 MeV when it is converted
into SQM, and the total energy released due to the collapse of the
whole crust can be as high as $10^{46}$ --- $10^{47}$\,\rm {erg}
\cite{2004Jia}, leading to an intense short explosion.
\cite{2018Zhang} argued that an FRB could be produced during the
crust collapse. However, if the SS has a strong dipolar magnetic
field, then the accreted material will fall toward the polar
region along the magnetic field lines and accumulates there. In
this case, only the polar region is overloaded and it may lead to
a local collapse rather than the collapse of the whole crust.
\cite{1998Cheng} suggested that such a local collapse can
interpret the quasi-periodic hard X-ray bursts from GRO J1744-28.
Very recently, \cite{2021Geng} also used the fractional collapse
of an SS crust to explain periodical repeating FRBs. In their
model, the SS accretes matter from a close-in companion star. The
accreted matter also accumulates at the two polar regions, leading
to a local collapse when the crust at the polar region is too
heavy. Note that the periodical behavior of FRBs is caused by the
thermal instability of the accretion disk, thus the repeating
period is not very strict. It is worth mentioning that the polar
cap is a surface region of the compact star determined by the open
magnetic field lines
\cite{1977Cheng,1978Scharlemann,1979Arons,1998Zhang}.

In this study, we consider the interaction between an SS and its
planet near the periastron. The planet is partially disrupted
there and the stripped material is accreted by the SS, falling to
the two polar cap regions and accumulated on the crust. It may
finally lead to a local collapse of the polar cap region,
producing an FRB. A schematic illustration of our model is
presented in Figure 1.

\begin{figure*}
\includegraphics[width=1\textwidth]{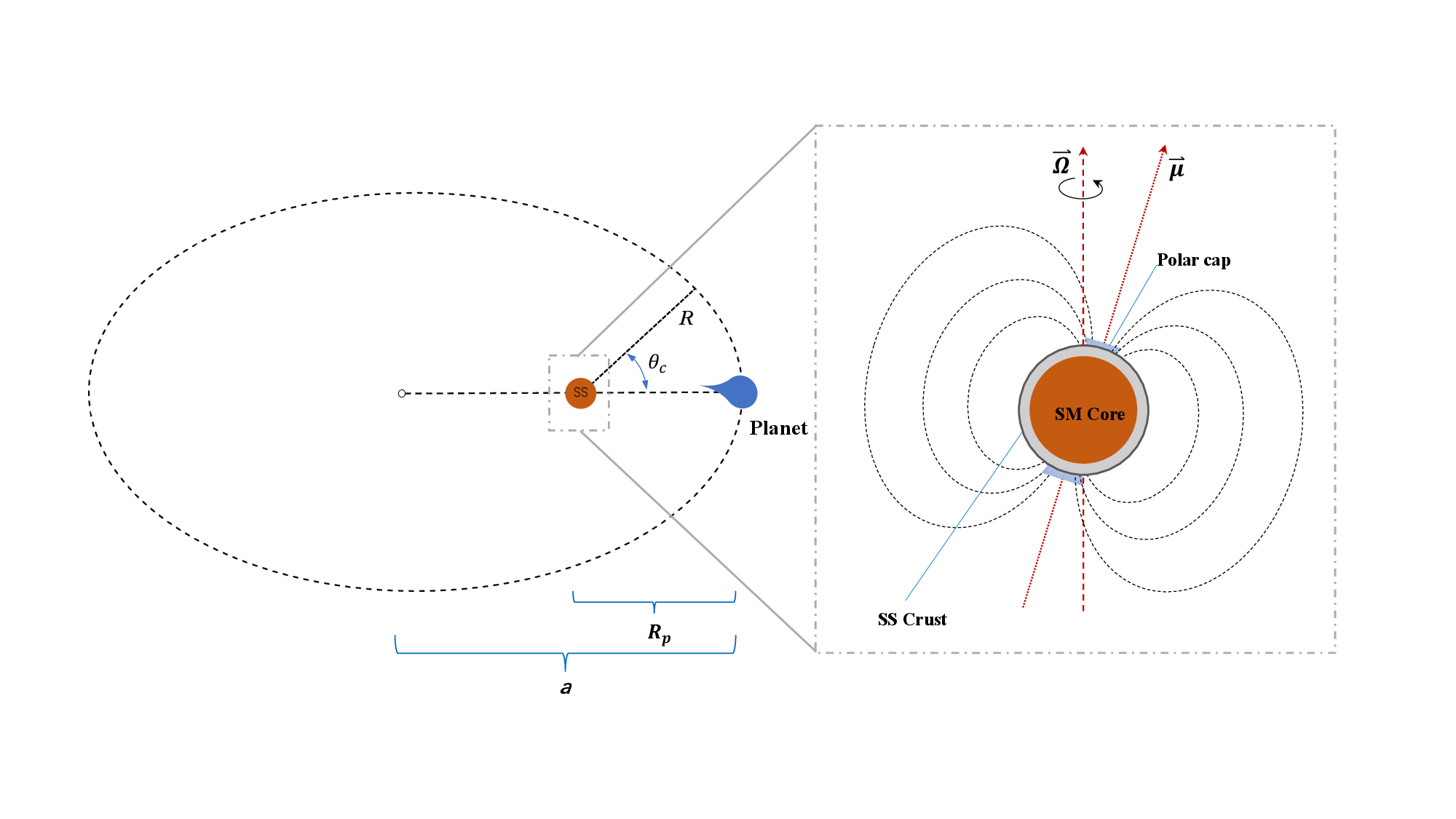}
\caption{A not-to-scale schematic illustration of our model. The
left portion shows that a planet is moving around an SS in a
highly eccentric orbit. It will be partially disrupted when
passing through the periastron. The right portion is an
enlargement of the SS. It shows that the accreted material
accumulates at the polar cap region of the SS, finally leading to
a local collapse of the crust.
Here $\vec{\mu}$ and $\vec{\Omega}$ indicate the
magnetic axis and rotation axis of the SS, respectively.}
\label{fig:fig1}
\end{figure*}

In our study, we take the SS mass, radius and total
crust mass as mentioned above, with a crust thickness of
$\ell \sim 2 \times 10^4\,\rm {cm}$. Then the mass of the crust at the
polar cap region is approximately
\begin{equation}
M_{\rm pl,crust} = \frac{\pi{\theta _{\rm cap}}^2}{4\pi} M_{\rm crust},
\end{equation}
where $\theta _{\rm {cap}}$ is the half opening angel of the polar cap.
It can be estimated by $\theta _{\rm {cap}} \approx(2 \pi r / c P)^{1/2}
\thicksim 1.45 \times 10 ^{-2} P_{0} ^{-1/2} r_{6}^{1/2} $, where
$P$ is the rotation period of the SS. $\theta _{\rm {cap}}$ implies a
filling factor of the surface of $f \approx \theta _{\rm {cap}}^{2} /4$
\cite{Kashiyama2013,2020Zhang}. Assuming each hadron can release
an energy of $\sim 6.3\,\rm {MeV}$ when it is converted to quark matter,
the total energy released during the local collapse of the polar
cap region is
\begin{equation}
E_{\rm {c,tot}} = \frac{M_{\rm {pl,crust}}}{m_{\rm p}} \times  6.3\,\rm {MeV},
\end{equation}
where $m_{\rm {p}}$ is the mass of proton.

The energy is released on a short timescale \cite{2004Jia},
leading to the formation of a fireball of electron/positron
($e^{\pm}$) pairs. The electrons/positrons streams outward along
the open magnetic field lines, producing a thin shell in the
magnetosphere. They will finally give birth to an FRB through
coherent emission \cite{2018Zhang}. Note that only a small
portion of the energy of $E_{\rm {c,tot}}$ will go into the FRB due to
the limited energy conversion efficiency in the process. Denoting
the energy conversion efficiency as $\eta$, then the intrinsic
energy of the FRB is $E_{\rm {FRB}} \sim \eta E_{\rm {c,tot}}$.

 After the collapse of the polar cap crust, the
stripped material from the planet will continue to fall toward
the polar cap region. The broken crust can then be re-built to
get ready for the next burst. Repeating FRBs can be produced
in this way as long as the accretion continues. Note that in each
collapse event, it is not necessary that all the matter of the
polar cap crust must fall onto the strange core. It is quite
possible that only a portion of the matter will do so that
significantly less material is needed to restore the crust. We
will further discuss this issue in the next section.

\section{comparison with FRB 180916}

FRB 180916 is a periodically repeating FRB source. Till the end of
2021, a total of 44 bursts have been detected from this source
\cite{2022Mckinven}. During the active state, an event rate of
about $0.9^{+0.5}_{-0.4}$ bursts per hour is reported. A period of $16.35 \pm 0.18$ days is found in
its activities, with a $\sim 5$ day active window in each period
\cite{CHIMEFRB:2020bcn}. Some plausible models have been
proposed to explain this periodic behavior of FRB 180916
\cite{LiQC2021,Sridhar2021}.

In this section, we use our model to explain the periodicity of
FRB 180916. First, we take typical values for some model parameters,
such as the SS mass and radius, and the total crust mass. 
For a typical spin period of $P = 1\,\rm s$, the crust mass at the
polar cap region will be $M_{\rm {pl,crust}} \approx 3.55 \times 10^{23}\,\rm {g} $.
The mass of the planet is taken as $ m_{\rm {planet}} = 10^{-3}\,M_{\odot}$
 and its mean density is taken as $\bar{\rho} = 10 \,\rm g\,cm^{-3}$.
 The radius of the planet is then $R_{\rm planet}
= \left(\frac{3 m_{\rm planet}}{4 \pi \bar{\rho}}\right)^{1/3}$.

In our framework, the observed period of 16.35 days in FRB 180916
is due to the orbital motion of the planet around the SS, i.e.
$P_{\rm orb} = 16.35 $ days. Then, from Equation (1) in Section 2, we
can derive the semimajor axis of the planet orbit as
$a=\left[G (M_{\rm ss}+ m_{\rm planet}) P_{\rm orb}^2/4\pi^2\right]^{1/3}
\simeq 2.11 \times 10^{12}\,\rm{cm}$. At the same time, the tidal disruption
radius is
 $R_ {\rm T} = \left(6M_{\rm ss}/ \pi \bar{\rho}\right)^{1/3} \simeq 8.08\times 10^{10},\rm {cm}$.
 So, from Equation (2), the orbital eccentricity is $e= 1 - 1.5 \,R_{\rm{T}}/a \simeq 0.94$ correspondingly.
 Note that we have taken the pericenter distance as $R_{\rm P} = 1.5 \,R_{\rm T} \simeq 1.21 \times  10^{11}$ \rm cm.
 For the clumps that are generated from the planet's inner side, the periastron distance can be written
 as $R_{\rm P}^{\rm cl} = a_{\rm cl}\left( 1-e_{\rm  cl}\right)$, where
 $a_{\rm cl} = a\left( 1+ a \frac{2R_{\rm planet}}{R_{\rm P}(R_{\rm P}+R_{\rm planet})}\right)$
 and $e_{\rm cl}= 1-\frac{R_{\rm P}-R_{\rm planet}}{a_{\rm cl}}$ are the
 semimajor axis and eccentricity of the clump, respectively \cite{2023MNRAS}.
 Then the clumps' periastron distance is
  $R_{\rm P}^{ \rm cl} \simeq 5.67\times 10^{10} \,\rm cm $.

 We assume that the SS has a very strong
   magnetic field of $ B = 1 \times 10^{16}\,\rm G $ and take the mass accretion rate
   as $\dot{m} = \dot{M}_{\rm Edd}$, then the magnetosphere radius of the SS can be calculated as
 $R_{\rm m} = \left( \frac{\mu^4}{2GM_{\rm ss}\dot{m}^2} \right)^{1/7} \simeq 3.09 \times 10^{10} \rm cm$,
 where $\mu = BR_{\rm ss}^3$ is the dipolar moment of the magnetic field. We see that the magnetosphere radius
 and the clumps' periastron distance is comparable, which indicates that
 the clumps would be seriously affected by the magnetic field. They will be quickly decelerated
 by the magnetosphere and be captured by the SS. In fact, if we consider a more compact
 planet with a higher density of $\bar{\rho} = 40\, \rm g\,cm^{-3}$ \cite{2013ApJS,2014ApJS},
 then the clumps' periastron distance is $R_{\rm P}^{\rm cl} = 2.75\times10^{10} \rm cm$.
 It is even smaller than the magnetosphere radius, so that the clumps will be
 effectively captured by the magnetic field and fall toward the polar cap region of the
 SS along the field lines.
 The falling timescale of the clumps (with a radius of $R_{\rm cl}$) from the magnetosphere
   radius to the vicinity of the SS can be roughly estimated as \cite{Perets}~
 $ t_{\rm fb} = \frac{2\pi R_{\rm m}^{3}}{\left(G M_{\rm ss}\right)^{1/2}\left(2R_{cl}\right)^{3/2}}$.
 For a clump of the size of 2 km, the falling timescale is $t_{\rm fb} \backsimeq 0.51\, \rm day$.
 Therefore, with the help of the magnetic field, the clumps could be essentially captured by the
 SS and fall toward the polar region very soon.

For FRB 180916, the observed duty cycle of bursts is about 5 days.
It corresponds to the duration of the partial disruption stage,
i.e. the time consumed by the planet when it passes through the
periastron. So, we have $\Delta P_{\rm {orb}} = 5$ days. Then, under
the Eddington accretion limit, the mass of the matter accreted by
the SS can be roughly estimated as $M_{\rm {acc}} = \Delta P_{\rm {orb}} \cdot
\dot{M}_{\rm {Edd}} \simeq 4.32 \times 10^{23}\,\rm {g} $. We see
that this mass is higher than the crust mass at the polar cap
region ($M_{\rm {pl,crust}}$), which means that the accreted material
will be sufficient to trigger at least one local collapse event.
Note that during one single local collapse process, not all the
material at the polar cap will necessarily fall onto the strange core.
It is quite possible that only a fraction of the matter will collapse. In this case,
an accreted mass significantly less than $M_{\rm {pl,crust}}$ will be
enough to restore the polar cap region. It means that more than one FRB
could be produced during one active cycle, which may account for the
relatively high burst rate of FRB 180916 in an active window. However,
it is very difficult to estimate the exact ratio of polar cap material
collapsed in the process, which needs further investigation in the future.

More interestingly, recent observations reveal that the
active window of FRB180916 can essentially be as large as one half of the period \cite{Bethapudi2023}.
Of course, the event rate at phases other than the normal 5-day duration is significantly
lower. This feature could also be explained in our framework.
In our model, the partial disruption happens in a 5-day duration. The disrupted materials
are then accreted by the SS. In realistic case, there should exist a
dispersion in the falling time of the materials. Also, the accretion rate at later
stages would be smaller, which also prolongs the accretion process.
Therefore, the total duty cycle will be correspondingly longer than 5 days and can extend
to one half of the period.

It should be noted that the periodicity of FRB 180916 is rather
stable ever since the discover of this repeating source \cite{2023Sand}. However, in our
model, the mass loss of the planet may lead to a change in its orbital period. We
thus need to examine this effect.
In an interacting binary system, the companion star will expand its radius to adjust
to a new equilibrium state after each partial disruption. The disruption
radius will change correspondingly, which leads the period to change.
For a normal companion star, such an effect has been studied in detail by
several groups \cite{1987Hjellming,2014Bogdanovi}. 
Here we follow Hamers \& Dosopoulou \cite{Hamers2019} to calculate the 
evolution of the orbital period of the planet. The detailed calculation procedure 
is described in the Appendix section. Our numerical results for the evolution of 
the orbital period is shown in Figure 2. We have considered four different mean 
densities for the planet, $\rho =1, 3, 8, 10 \,\rm g\,cm^{-3}$.
Generally, we see that the period is relatively stable at early stages in all the cases.
Especially, the period change is very small in 200 --- 400 years.
Also, the total mass loss of the planet is very small comparing to its original mass
during this time. Thus our model is consistent with the stable periodicity observed in FRB 180916.

\begin{figure*}
  \includegraphics[width=1\textwidth]{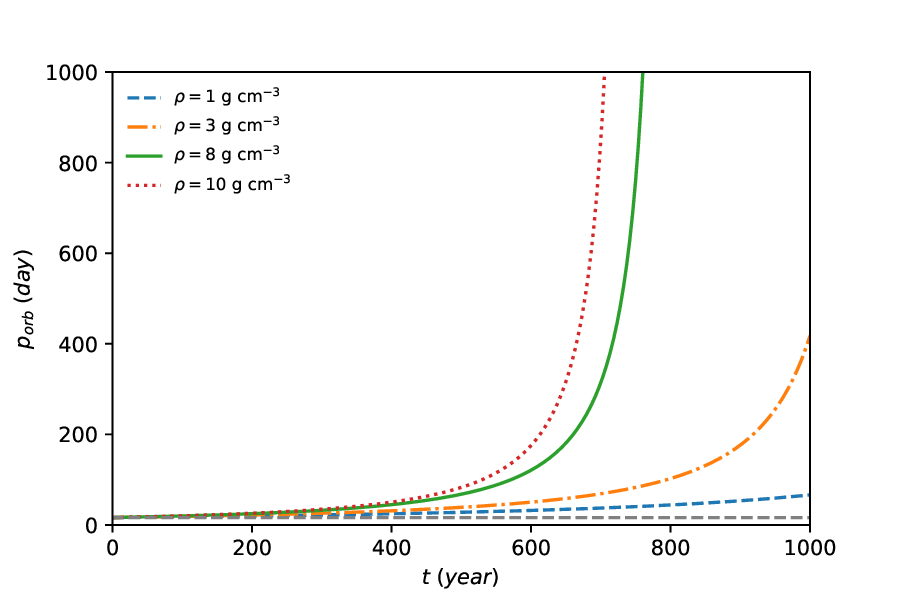}
  \caption{ Long-term evolution of the orbital period of the planet.
  We take the parameters as $m_{\rm planet} = 10^{-3}\,\rm M_{\odot}$,
  $M_{\rm ss}= 1.4\,\rm M_{\odot }$, and $R_{\rm p} = 1.5 R_{T}$.
  The initial period is taken as $P_{\rm orb} = 16.35$ days, which is shown by
  the dashed horizontal line. Different mean densities are assumed for the
  planet, which are marked in the figure. }  
  \label{fig:fig1}
  \end{figure*}

Taking $M_{\rm {pl,crust}} \approx 3.55 \times 10^{23}\,\rm {g} $, the
collapse of the polar cap region will release a total energy
of $E_{\rm {c,tot}}\approx 2.15 \times 10^{42}\, \rm erg$ in one
orbital period. The energy of FRBs
can then be $E_{\rm {FRB}} \sim  2.15 \times 10^{38}\,\rm erg$
even for a very small conversion efficiency of $\eta = 10^{-4}$. It
is large enough to account for most observed bursts from FRB 180916.
Furthermore, if the FRB emission is highly beamed, then the energy will
be emitted into a very narrow solid angle. It means that the observer would
see a much stronger burst (with a large isotropic-equivalent energy).
Since the emissions of FRBs are coherent and strong
magnetic fields are involved, the radiation should naturally be beamed,
which significantly amplifies the isotropic-equivalent energy.  
%% In other words, if the FRB was effected by a beaming effect,
%% then its isotropic energy would be even much higher.

Recent observations indicate that the
active window of FRB 180916 seems to be
frequency-dependent\,\cite{2021Natur,2021Pleunis,LiQC2021}.
FRB emissions have been detected from 1.4 GHz down to 120 MHz.
In such a wide frequency range, the active window is chromatic,
i.e., bursts with higher frequencies generally arrive earlier
in phase. Such a frequency-dependent behavior could be
explained in our framework. As described earlier in this section,
the observed duty cycle of bursts corresponds to the duration of
the partial disruption stage, i.e. the time taken for the planet
to pass by the periastron. In this period, the stripped material
from the planet will gather around the SS, producing a plasma
cloud surrounding the compact star.
 The density of the plasma will decrease with
time due to the accretion of the SS, which will lead to a decrease
of the characteristic oscillation frequency of the plasma.
As a result, low frequency FRBs cannot penetrate through the
plasma at early stages of the active period. They are detectable
only when the plasma's characteristic oscillation frequency becomes
small enough at relatively late stages of the duty cycle.

To present a more detailed explanation of the
process, let us assume that the accretion of the clump debris by
the SS follows the normal $t^{-5/3}$ rule of tidal disruption
events~\cite{1988Rees,2023Kaur}. Then, the accretion rate varies
with time as $\dot{m}\sim \dot{M}_{\rm fb}\left(\frac{t}{t_{\rm fb}} \right)^{-5/3} $,
where $\dot{M}_{\rm fb}$ is the accretion rate at the fallback time ($t_{\rm fb}$).
Consequently, the plasma density around the SS also decays
as $n_{\rm e} \sim n_{\rm e0 } \left(\frac{t}{t_{\rm fb}} \right)^{-5/3}$,
where $n_{\rm e0 }$ is the density at $t_{\rm fb}$. The characteristic
oscillation frequency of the plasma can then be calculated
as $\omega_{\rm p} =\left( \frac{4\pi n_{\rm e}e^{2}}{m_{\rm e}}\right)^{1/2}
\simeq \left(5.63 \times 10^{4} \rm \,s^{-1}\right)n_{\rm e}^{1/2}$, where
$n_{\rm e}$ is in units of cm$^{-3}$ in the last expression.
Assuming that most of the material falls onto the SS to produce the FRBs
and only a fraction of 10\% still remains in the magnetosphere initially,
then the density of the remnant plasma can be estimated as $n_{\rm e0} =
\frac{3m_{\rm cl}}{40 \pi R_{\rm m}^{3}m_{\rm P}}$. For typical clumps
with $R_{\rm cl} = 2\, \rm {km} $ and $m_{\rm cl} = 1.34\times 10^{18}\rm g$,
the initial plasma density is $n_{\rm e0}  = 6.49 \times 10^{8}\,\rm cm^{-3}$
and the corresponding characteristic frequency
is $\omega_{\rm p} = 1.43 \times 10^{9}\,\rm Hz $. Taking $t_{\rm fb} = 0.5$ days,
the plasma density will decrease to $n_{\rm e} = 6.66 \times 10^7 \,\rm cm^{-3}$
two days later, and the corresponding characteristic oscillation frequency
is $0.46 \times 10^{9}$\,Hz. We see that the plasma frequency decreases
significantly during the duty cycle. It means that high frequency FRBs
could penetrate through the magnetosphere and would be detected at
the early stage of the duty cycle. However, low frequency events will
be absorbed by the plasma at early stages. They could be detected
mainly at later stages, when the plasma density becomes smaller. Such a
picture is well consistent with the frequency-dependent active window
as observed in FRB 180916.

Note that there a small random variation of
the dispersion measure (DM) observed in FRB 180916. Its
amplitude is of the order of $\sim 1\,\mathrm {pc}\, \mathrm {cm}^{-3}$.
In our model, additional electrons will be emitted from the polar
cap region when the crust collapses to produce an FRB, the mean
density of which can be estimated as
$n_{\rm e}= \frac{ \zeta E_{\mathrm{c, tot}}}{\delta t}
\frac{1}{4\pi m_{\mathrm p}c^{3}R_{\mathrm FB}^{2} 2\Gamma^{2}} $,
where $\delta t $ is the burst duration, $\mathrm{R_{\mathrm FB}}$
is the typical radius of the electron cloud,
$\eta$ is the radiation efficiency, and $\Gamma$ is the
Lorentz factor of electrons.
Taking $\eta \sim 10^{-4}$, $\delta t \sim 10 \,\mathrm {ms} $,
$\Gamma = 100$, and $R_{\rm FB} \sim 1 \times 10^{8}$ cm,
the mean electron density of the fireball will
be $n_{\rm e} \sim 6.4 \times 10^{10}\,\mathrm {cm}^{-3}$.
It will lead to an additional $DM$ of $\sim 2\,\mathrm{pc}\,\mathrm{cm}^{-3}$,
which is roughly consistent with the variation amplitude of the observed $\rm{DM}$.

Furthermore, for FRB 180916,
the subpulses are found to show a downward drifting behavior on the
time-frequency diagram. It could be easily explained by considering
the radius-to-frequency mapping in our framework. In our scenario,
the radio emission comes from the coherent curvature radiation of
pairs along open magnetic field lines, then the later subpulses
usually have a lower characteristic frequency since
they are emitted at higher latitudes \cite{WangWY2022}. It
is quite similar to the radio emission of pulsars.

Observations show that FRB 180916 has an obvious secular increase
in the rotation measure ($\Delta RM \sim 50$ rad/m$^{-2}$) over the 9-month period
from 2021 April to 2021 December \cite{2023Mckinven}. Long-term rotation measure (RM)
variabilities have been explained in the framework of massive binary systems
\cite{Wang2022,Zhao2023}. However,
the secular RM variability of FRB 180916 is not associated with any significant DM variability,
nor does it be connected with the periodicity and active window variations \cite{2023Sand,2023Mckinven}.
Such a RM variation thus is not a natural outcome of binary interactions, 
including our model. We suggest that it could be
caused by some environmental effects. It is possible that there are some inhomogeneous
magnetic fields around the FRB source, which lead to the change of RM as the line of 
sight changes due to the proper motion of the SS. 

\section{Conclusions and discussion}

\label{sec:conclusion}

In this study, a new model is proposed to explain the periodical
behavior of FRB 180916. We consider the interaction between an SS
and a planet. The planet is assumed to move around the SS in a
highly eccentric orbit, with an eccentricity of $e = 0.94$. The
orbital period is 16.35 days, which just corresponds to the
observed period of FRB 180916. The periastron of the planet
satisfies the condition of $R_{\rm {T}} < R_{\rm {p}} < 2.7 R_{\rm {T}}$, so that it will
be partially disrupted by tidal force of the SS every time it
passes through the periastron. The duration of such a tidal
interaction near the periastron is about 5 days in each period,
which explains the observed duty cycle in FRB 180916.
The stripped material from the planet is then
captured and accreted by the SS due to the strong magnetic field
of the compact star, flowing to the polar cap region along the
field lines and accumulated there.
The total mass of the accreted material is estimated
as $M_{\rm {acc}} \simeq 4.32 \times 10^{23}\,\rm {g}$ in each period, large
enough to lead to local collapses of the crust at the polar cap
region. A total energy of $E_{\rm {c,tot}} \approx 2.15 \times 10^{42}\,\rm
{erg}$ will be released during the collapse. Assuming a small
conversion efficiency of $\eta = 10^{-4}$, it is still sufficient
to account for typical FRBs.

In our framework, the magnetosphere of the SS is
filled by a small portion of the disrupted material. The resultant
plasma density will gradually decrease as the accretion proceeds,
leading to a decay in the characteristic oscillation frequency of
the plasma. As a result, high frequency FRBs would be observed
mainly in the early phase of the duty cycle, while low frequency
events are observable generally at later stages. It naturally
explains the frequency-dependent active window of FRB 180916.
Each time a collapse occurs, a fireball consisted of a large
number of electrons/positrons will be released, leading to an
additional DM variation of $\sim 2\,\rm pc\, cm^{-3}$. It is also
roughly consistent with the observed DM variation of FRB180916.

Note that \cite{2021Geng} also tried to explain the periodically
repeating FRBs by adopting the local collapse of the crust of an
SS at the polar cap region. The difference in their model is that
the accreted material is supplied by a companion star at a
constant rate. The periodical behavior is then due to
thermal-viscous instability of the accretion disk. In their
framework, the period and the active window are determined by the
accretion rate and the viscosity parameter. A natural result is
that the period in their model may not be stable, but may vary in
a relatively wide range. On the contrary, according to our
modelling in this study, the FRB period is simply the orbital
period of the planet, which is thus much more rigorous.

The interaction of planets/asteroids with pulsars has previously
been suggested as the mechanism for producing FRBs by many authors
\cite{2015Geng,2020Dai,Geng_2020,2021Voisin,2022Kurban}.
Especially, \cite{2022Kurban} studied the interaction between an
NS and its planet in an elliptical orbit, where the partial
disruption of the planet is also considered. The difference
between those models and our current model is significant.
Generally, in those models, the compact star is an NS, but not an
SS. As a result, the energy mainly comes from the gravitational
potential energy of the planets/asteroids (or from the
Alfv$\acute{\rm e}$n wing interaction in \cite{2022Kurban}'s model).
On the contrary, in our current model, the energy mainly comes from
the de-confinement of normal hadronic matter at the polar cap region.
Furthermore, comparing with previous
studies \cite{2018Zhang,2021Geng,2022Kurban}, here we have gone
further to present more details in the modeling. The frequency-dependent
active window, the variation of dispersion measure and the downward
drifting behavior in the frequency domain observed in FRB 180916 can all be
naturally explained in our framework. Thus our modeling is a
useful supplementary to the previous studies.

In our modeling, we have mentioned that it is quite
possible that only a fraction of the polar cap crust
will fall onto the SQM core during a collapse event. The energy
released is enough to produce an FRB, especially
when the radiation is highly beamed. An advantage in this case is
that less material is needed to re-build the polar cap crust
again. It means that more than one FRB could be produced in the
duty cycle of one period. Anyway, the issue that what fraction of
the polar cap crust will be destroyed during the collapse is a
complicated problem and is beyond the scope of this study. It
deserves further study in the future.

Recently, \cite{2022Li} argued that an energetic fireball of
electron/positron pairs could be erupted when a crack forms and
then heals during a starquake event in neutron stars. The
electrons and positrons can then produce some kinds of transients
or bursts through their interaction with the magnetosphere. It is
interesting to note that the collapse of the polar cap crust in
our framework is somewhat like a starquake event. Similarly, an
outburst might also be produced in other wavelengths, such as in
X-rays and optical wavelength. However, such an outburst might be
detected only when the source is not too far from us.

\section{Acknowledgments}
We are grateful to the anonymous referee for valuable comments and
suggestions that lead to an overall improvement of this study.
We thank Chen-Ran Hu and Chen Deng for useful discussion.
This study was supported by the National Natural Science Foundation
of China (Grant Nos. 12233002, 12041306, U1938201, 12273113, 12103055),
by the Major Science and Technology Program of Xinjiang Uygur Autonomous Region (No. 2022A03013-1),
by National SKA Program of China No. 2020SKA0120300, by the National Key R\&D
Program of China (2021YFA0718500).
YFH also acknowledges the support from the Xinjiang Tianchi Program.

\appendix
\section{Evolution of the orbital period of a close-in binary system}

In our framework, the radius of the planet will expand to adjust
to a new equilibrium state after each partial disruption. The disruption
radius will change correspondingly. As a result, the orbital period will
also change. In our study, we follow Hamers \& Dosopoulou \cite{Hamers2019} 
to calculate the evolution of the orbital period. 
The method is particularly suitable for solving the secular 
(i.e., orbit-averaged) changes of the orbital elements due to mass transfer
in eccentric binaries. 
	
For an interacting eccentric binary system, Hamers \& 
Dosopoulou \cite{Hamers2019} considered 
two types of mass transfer: (i) The donor fills its Roche lobe 
during the entire orbit so that Roche lobe overflow (RLOF) takes place all 
the time. This is called the full RLOF case, which happens if the binary 
is too close-in; (ii) The donor fills its Roche lobe only during part 
of the orbit. It happens when the apastron is large but the orbit 
is very eccentric. This is called the partial RLOF case. The condition 
considered in our model obviously fits to their partial RLOF case. 

The donor moves in an eccentric orbit defined by 
\begin{align}
	r(\theta) = a(1-e\cos \theta),
\end{align}
where $\theta$ is the true anomaly. Partial RLOF takes place 
in an orbital phase range of $-\theta_{\rm c} < \theta < + \theta_{\rm c}$. 
Note that the critical phase $\theta_{\rm c}$ has been defined just below 
Equation (2) in the main text. 
Ignoring the spin of the donor star and assuming that the mass transfer 
is conservative, then the orbit-averaged dynamical equations can be 
expressed as \cite{Hamers2019} 
\begin{align}\label{eq:6}
  \displaystyle\frac{\langle \dot{a} \rangle}{a} &=\displaystyle -\frac{2\langle \dot{M_d} \rangle}{M_d} \frac{1}{f_{\dot{M}}(e,x) } \biggl[ (1-q) f_a(e,x) + X_{L,0}(q) g_a(e,x)  - q \frac{r_{A_a}}{a} h_a(e,x) \biggr],\\ \label{eq:7}
  \displaystyle \langle \dot{e} \rangle &= \displaystyle -\frac{2 \langle \dot{M_d} \rangle}{M_d} \frac{1}{f_{\dot{M}}(e,x) } \biggl[ (1-q) f_e(e,x) + X_{L,0}(q) g_e(e,x)  - q \frac{r_{A_a}}{a} h_e(e,x) \biggr], 
  \end{align}
where $q$ is the mass ratio of the two objects, $M_{\rm d}$ is the mass of the 
donor star and $\dot{M_{\rm d}}$ is its mass loss rate.  
$r_{\rm {A_a}}$ is the separation between the initial position of the accreted 
matter and the accreting star.  
$x$ is the ratio of the Roche lobe radius in a circular 
orbit ($R_{\rm L}^{\rm c}$) to the radius of the donor,  
\begin{align}
	\label{eq:x_def}
	x \equiv  \frac{R_\mathrm{L}^\mathrm{c}}{R} = \frac{a}{R} \frac{0.49 \, q^{2/3}}{0.6 \, q^{2/3} + \ln\left (1 + q^{1/3} \right )}. 
\end{align}
$X_{L,0}(q)$ is the normalized position of the first Lagrange 
point, which can be analytically expressed as \cite{Hamers2019}
\begin{align}
	X_{L,0}(q) = \frac{1}{6} \left(-\sqrt{3} \sqrt{\frac{6 \sqrt{3}
			(q+1)}{\sqrt{A_{-}+A_{+}-2 q+3}}-A_{+}-\frac{q^2}{A_{+}}-4 q+6}+\sqrt{3}
	\sqrt{A_{-}+A_{+}-2 q+3}+3\right), 
\end{align}
with 
\begin{align}
	A_{\pm} \equiv \sqrt[3]{q \left(q^2\pm6 \sqrt{3} \sqrt{q^2+27}+54\right)}.
\end{align}

The definition and detailed expressions of other functions in 
Equations (\ref{eq:6}) and (\ref{eq:7}), such as $f_{\dot{M}}(e,x)$, 
$f_a(e,x)$, $g_a(e,x)$, $h_a(e,x)$, $f_e(e,x)$, $g_e(e,x)$ 
and $h_e(e,x)$, are given below \cite{Hamers2019} : 
\begin{align}
    \label{eq:f_M}
    \begin{split}
     f_{\dot{M}} (e,x) &=-\frac{1}{96 \pi } \Biggl [ 36 e^4 \theta_0 x^3+3 e^4 x^3 \sin (4 \theta_0)-32 e^3 x^3 \sin (3 \theta_0)+24 e^3 x^2 \sin (3 \theta_0)+288 e^2 \theta_0 x^3-432 e^2
       \theta_0 x^2\\
    &\quad+24 e^2 x \left(\left(e^2+6\right) x^2-9 x+3\right) \sin (2 \theta_0)-24 e \left(4 \left(3 e^2+4\right) x^3-9 \left(e^2+4\right)
       x^2+24 x-4\right) \sin (\theta_0)\\
       &\quad+144 e^2 \theta_0 x \quad+96 \theta_0 x^3-288 \theta_0 x^2+288 \theta_0 x-96 \theta_0\Biggl ], 
    \end{split}
\end{align}
\begin{align}
      \begin{split}
      f_a(e,x) &= \frac{1}{96 \pi } \Biggl [ 36 e^4 \theta_0 x^3+3 e^4 x^3 \sin (4 \theta_0)-16 e^3 x^3 \sin (3 \theta_0)+24 e^3 x^2 \sin (3 \theta_0)-144 e^2 \theta_0 x^2\\
      &+24 e^2 x\left(e^2 x^2-3 x+3\right) \sin (2 \theta_0)-24 e \left(\left(6 e^2-8\right) x^3+\left(12-9 e^2\right) x^2-4\right) \sin (\theta_0)+144 e^2\theta_0 x\\
      &-96 \theta_0 x^3+288 \theta_0 x^2-288 \theta_0 x+96 \theta_0 \Biggl ], 
      \end{split}
\end{align}
\begin{align}
  \begin{split}
  g_a(e,x) &= \frac{1}{32 \pi } \Biggl [ 4 \theta_0 x \left(e^2 \left(\left(e^2-8\right) x^2+12\right)-8 ((x-3) x+3)\right)\\
  &+e x \Biggl \{e \biggl [8 \left(x \left(\left(e^2+2\right)
  x-6\right)+3\right) \sin (2 \theta_0)+e x (3 e x \sin (4 \theta_0)\quad-16 (x-1) \sin (3 \theta_0))\biggl ]\\
  &-16 x \left(e^2 (x-3)-4 x+6\right) \sin(\theta_0)\Biggl \} \Biggl ], 
  \end{split}
\end{align}
\begin{align}
  \begin{split}
  h_a(e,x) &= \frac{1}{4 \pi } \Biggl [ e \sin (\theta_0) \left(-\left(e^2-4\right) x^3-\frac{4}{e \cos (\theta_0)-1}-12 x\right)+x \Bigl \{ e^2 (-x) (e x \sin (3 \theta_0)\\
  &-3 (x-2)\sin (2 \theta_0))-2 \theta_0 \left(x \left(\left(e^2+2\right) x-6\right)+6\right)\Biggl \}+\frac{8}{\sqrt{1-e^2}} \tan ^{-1}\left(\sqrt{\frac{1+e}{1-e}} \tan \left(\frac{\theta_0}{2}\right)\right) \Biggl ], 
  \end{split}
\end{align}
\begin{align}
	\begin{split}
		f_e(e,x) &= \frac{1-e^2}{32 \pi } \Biggl[e x \left(12 e^2 \theta_0 x^2+e^2 x^2 \sin (4 \theta_0)+8e x \left(\left(e^2+3\right) x^2-6 x+3\right) \sin (2\theta_0)\right)\\
		&-e x \left(8 e (x-1) x \sin (3 \theta_0)+48 \theta_0 x^2-96 \theta_0 x+48 \theta_0\right)-8 (x-1) \left(\left(9 e^2+4\right) x^2-8 x+4\right)\sin (\theta_0)\Biggl], 
	\end{split}
\end{align}
\begin{align}
  \begin{split}
  g_e(e,x) &= \frac{1-e^2}{48 \pi  e} \Biggl [12 \theta_0 \left(e^2 x \left(x \left(\left(e^2+4\right) x-9\right)+6\right)-2\right)+e^2 x \Biggl \{ 6 \left(x \left(2\left(e^2+4\right) x-15\right)+6\right) \sin (2 \theta_0)\\
  &+e x (3 e x \sin (4 \theta_0)+(18-20 x) \sin (3 \theta_0))\Biggl \}-6 e \left(x
  \left(e^2 x (14 x-15)+8 (x-3) x+24\right)-4\right) \sin (\theta_0)\\
  &+48 \sqrt{1-e^2} \tan ^{-1}\left(\sqrt{\frac{1+e}{1-e}} \tan \left(\frac{\theta_0}{2}\right)\right)\Biggl ], 
  \end{split}
\end{align}
\begin{align}
  \begin{split}
  h_e(e,x) &= \frac{1}{48 \pi  e} \Biggl [ \frac{e^2-1}{e \cos (\theta_0)-1} \Biggl \{
  16 e^4 x^3 \sin (2 \theta_0)+4 e^4 x^3 \sin (4 \theta_0)-26 e^3 x^3 \sin (3 \theta_0)+27 e^3 x^2 \sin (3 \theta_0)\\
  &+24 e^2 \theta_0x^3+60 e^2 x^3 \sin (2 \theta_0)-36 e^2 \theta_0 x^2-126 e^2 x^2 \sin (2 \theta_0)-12 e \theta_0 \left(e^2 x^2 (2 x-3)-2\right) \cos(\theta_0)\\
  &-3 e \biggl [2 \left(7 e^2+8\right) x^3-3 \left(3 e^2+16\right) x^2+48 x-8\biggl ] \sin (\theta_0)+72 e^2 x \sin (2 \theta_0)-24\theta_0 \Biggl \}\\
  &+144 \left(1-e^2\right)^{3/2} x \tan ^{-1}\left(\sqrt{\frac{1+e}{1-e}} \tan \left(\frac{\theta_0}{2}\right)\right)+48 \sqrt{1-e^2} \left(3\left(e^2-1\right) x+1\right) \tan ^{-1}\left(\sqrt{\frac{1+e}{1-e}} \tan \left(\frac{\theta_0}{2}\right)\right) \Biggl].
  \end{split}
\end{align}

The above expressions have been implemented by Hamers \& 
Dosopoulou into an easy-to-use PYTHON code \cite{Hamers2019}, 
which is freely available at 
gethub \footnote{\url{https://github.com/hamers/emt}}.
Adopting their code and considering the conditions in our 
framework, we have calculated the evolution of the orbital 
period of the planet. The main parameters involved are taken 
as: $M_{\rm d} = M_{\rm planet} = 10^{-3}\,M_{\odot}$,
$M_{\rm ss} = 1.4 \,M_{\odot}$, $a=0.1\,\rm au$, $e=0.94$. 
Our numerical results are presented in Figure 2 in the main text.
%% $\langle \dot{M_d} \rangle= -10^{-8}\,M_{\odot}\,\mathrm{yr^{-1}}$,
\bibliographystyle{apsrev4-1}
\bibliography{reference.bib}
\end{document}